\documentclass[11pt]{article}
\pdfoutput=1

\usepackage{natbib}
\usepackage{float}
\usepackage[a4paper, total={6.5in, 8in}]{geometry}
\usepackage{graphicx}
\usepackage{booktabs}
\usepackage[caption = false]{subfig}

\begin{document}
	
\begin{titlepage}
	\title{Donald Duck Holiday Game: A numerical analysis of a\\ Game of the Goose role-playing variant}
	\author{W.J.A. van Heeswijk \thanks{University of Twente, Department of Industrial Engineering \& Business Information Systems}}
	\date{\today}
	
	\maketitle

	\begin{abstract}
	The 1996 Donald Duck Holiday Game is a role-playing variant of the historical Game of the Goose, involving characters with unique attributes, event squares, and random event cards. The objective of the game is to reach the camping before any other player does. We develop a Monte Carlo simulation model that automatically plays the game and enables analyzing its key characteristics. 
	
	We assess the game on various metrics relevant to each playability. Numerical analysis shows that, on average, the game takes between 69 and 123 rounds to complete, depending on the number of players. However, durations over one hour (translated to human play time) occur over 25\% of the games, which might reduce the quality of the gaming experience. Furthermore, we show that two characters are about 30\% likely to win than the other three, primarily due to being exposed to fewer random events.  We argue that the richer narrative of role-playing games may extend the duration for which the game remains enjoyable, such that the metrics cannot directly be compared to those of the traditional Game-of-the-Goose.
	
	Based on our analysis, we provide several suggestions to improve the game balance with only slight modifications. In a broader sense, we demonstrate that a basic Monte Carlo simulation suffices to analyze Game-of-the-Goose role-playing variants, verify how they score on criteria that contribute to an enjoyable game, and detect possible anomalies.
	\end{abstract}
\end{titlepage}

\section{Introduction}\label{sec:introduction}
In the summer of 1996, the Donald Duck Holiday Game (Dutch: \textit{Donald Duck's Vakantiespel}, we will use the abbreviation DDHG) was attached as a board game to three issues of the Dutch magazine \textit{Donald Duck Weekblad} \citep{degeillustreerdepers1996}. It is a variant of the historical Game of the Goose (GG) board game, which history dates back as far as the 16th century \citep{seville2009}. It is documented that Francesco I de 'Medici of Florence -- the grand-duke of Tuscany from 1574 to 1587 -- gifted a version of the game to King Philip II of Spain, and that John Wolfe formally registered the game in England in 1597 \citep{storrier2006,duggan2016}.

The mechanism behind this spiral race game is exceedingly simple. The objective is to reach the final square on the board before any other player, while avoiding obstacles that hamper progression. The game is turn-based with a fixed player sequence. The active player casts a die to determine the number of squares the board character moves forwards. When encountering an event square, the player follows the instructions corresponding to the event. Thus, the game's progression is determined solely by randomness, not requiring any skill or proficiency on the player's part. The main entertainment value is rooted in malicious pleasure, watching with glee as other players are stuck in a well or thrown back to their starting position. However, as \citet{seville2009} notes, this thrill of uncertainty is meant to last for a relatively short time. From personal experience of the author, players in the DDHG are more than occasionally exposed to the whims of random events longer than necessary. This paper sets out to explore whether the sometimes excessive duration is merely a figment of imagination, or the game is actually unfairly cruel at times. At the same time, we reflect on possible reasons that keep the game enjoyable for a greater number of rounds than the original GG.

The traditional game of the goose consists of 63 squares, the DDHG considerably expands by offering 115 squares and distinct characters, as well as random event cards drawn from a shuffled deck. Following the classification of \citep{seville2009}, it would be a role-playing variant of the original game. The objective of the DDHG is to reach the camping located at the final square, ensuring to finish exactly on the spot. As a preceding square triggers an event that sends the player all the way back to start, traumatizing moments occur frequently near the end. Although the DDHG was eminently designed to keep children occupied for some time during the summer holidays and grant their parents a well-deserved break, one might argue that the game requires more effort, will-power and dedication than could reasonably be expected, from young children and adults alike. The sometimes markedly slow progress of the game may culminate into various stages of boredom, frustration, and eventually even violating the sacred game rules to hasten victory.

An apparent problem is that the ratio between positive and negative events is strongly skewed towards the negative. Only few events may be viewed as positive as they (potentially) propel the player forward, most of them being curses in disguise as they trigger picking up another random event card (which is likely to bear negative consequences). By developing a simulation model that enables repeated gameplay, this paper aims to verify whether excessive duration of the game is a matter of perception or that concerns about overly lengthy gameplay are justified, addressing both the average duration and the excesses. 

Another long-standing matter addressed is that of strategic character selection (although the game rules dictate that characters are randomly assigned by a die roll). The author usually chose the characters of Huey, Dewey \& Louie, bravely traveling the vast distance to the camping afoot, whereas all other characters utilize some vehicle during their journey. Personal bias aside, the random event cards in the game seem to mildly favor these seemingly disadvantaged characters. We investigate whether certain characters have a larger likelihood to emerge victorious. The game's description explicitly state that each character has an equal chance of winning, this paper attempts to either confirm or disprove that claim. In addition, we hope that the findings irrefutably prove that non-motorized transport modes outperform motorized vehicles, serving as potentially valuable evidence to support the author's research in sustainable transport.

This paper relates to several other works on the GG and its variants. \citet{seville2001} and \citet{neto2016} describe the use of Monte Carlo simulation to perform numerical experiments on the game, including distributional properties of the game duration. \citet{neto2016} explicitly focus on quantifying criteria that contribute to the games dramatic properties (as a certain exposure to drama keeps the race interesting), providing metrics on what constitutes an enjoyable game. We believe that the role-playing dimension studied in this paper makes for an interesting addition to existing works.

The contribution of this paper is as follows. We evaluate the DDHG on various criteria that contribute to enjoyable role-playing variants of the Game-of-the-Goose, paying attention to both quantitative and qualitative aspects. Based on our analysis, we show that some simple modifications considerably improve the balance of the game.

\section{Game description}\label{sec:description}
\begin{figure}[H]
	\includegraphics[width=\textwidth]{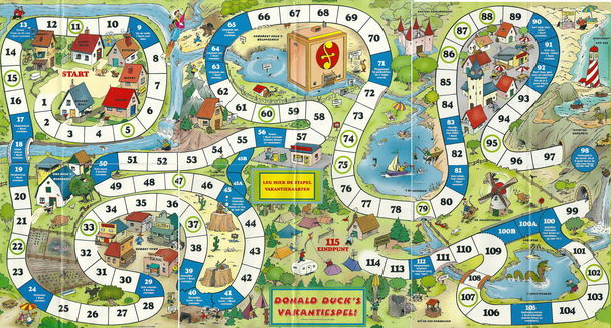}
	\caption{Image of \textit{Donald Duck's Vakantiespel}. Blue squares mark events, squares with green circle require drawing a random event cards. \copyright De Ge\"illustreerde Pers, 1996} \label{fig:donaldduckholidaygame}
\end{figure}

The game's objective is to be the first player to reach the camping located at the final square, but it is a road filled with a myriad of obstacles, challenges and experiences that must be endured before enjoying the rightfully deserved holiday. Figure~\ref{fig:donaldduckholidaygame} shows the game board.

The players are set in a fixed sequence. First, the players must throw a die to determine which character's mantle they assume; the characters are introduced in Section~\ref{ssec:characters}. To set the game into motion, one player must throw 6 to be the first to start. Subsequently, each player simply moves forwards according to the number thrown with the die (unlike other game variants, a value of 6 does not merit another throw). If the player lands on an event square (Section~\ref{ssec:randomeventsquares}) or a square that requires drawing a random event card (Section~\ref{ssec:randomeventcards}), the corresponding instructions of the event are followed. Typically, this involves waiting a number of turns or moving to another square, sometimes depending on the character attributes.

The game is only finished when a player ends exactly at Square $[115]$ (we use the notation $[\,\,]$ when referring to board squares). Any excess number of steps resulting from the die throw must be traversed backwards. As Square $[112]$ triggers an event that sends the player all the way back to start, there is a certain anxiety involved in attempting to secure the finish.

As a small touch of fortune, there are two shortcuts incorporated in the game. A bicycle lane connects Square $[45]$ to $[55]$ (only accessible on foot or by bicycle, when landing exactly on Square $[45]$), whereas the highway connecting Square $[100]$ to $[109]$ may only be entered by motorized vehicles and only when landing on Square $[100]$. Each shortcut has a length of two squares. We assume that, when allowed, the player will always take the shortcut. The only exception is when taking the shortcut would land the player on Square $[112]$ and subsequently sent back to start; in that case it is naturally wiser to take the detour.

\subsection{Characters}\label{ssec:characters}
There are five playable characters in the game, each with a unique mode of transport: Huey, Dewey \& Louie (Walking, under the proud banner of the Junior Woodchucks), Clarabelle (Bicycle), Horace (Motorcycle), Goofy (Bus), and Donald (Car; to be more precise, the infamous \textit{Duckatti 313}). As stated in the introduction, the game description asserts that each character has an equal chance of winning. From a gaming perspective, the car and bus are equivalent transport modes. It might seem odd that Donald chooses to vacation without his nephews and that the couple Horace and Clarabelle travels separately to their holiday destination, but such concerns are outside the scope of this paper. The character attributes are summarized in Table~\ref{table:characters}.

\begin{table}[H]
		\caption{Description of characters.}
		\label{table:characters}
			\small
\begin{tabular}{ l  l l}
	\textbf{Character} & \textbf{Transport mode} & \textbf{Notes} \\
	\toprule			
	Huey, Dewey \& Louie & Walking & \textbullet\ Affected by rain and wind \\
	&&	\textbullet\ Suspect to blistering\\
	&&	\textbullet\ May use bicycle lane\\
	Clarabelle & Bicycle & \textbullet\ Affected by rain and wind \\
&&	\textbullet\ Risk of flat tires\\
&&	\textbullet\ May use bicycle lane\\
	Horace & Motorcycle & \textbullet\ Affected by rain and wind \\
	&&	\textbullet\ Risk of flat tires \\
	&  & \textbullet\ Prone to engine failure \\
	&&	\textbullet\ May use highway\\
	Goofy & Bus & \textbullet\ Risk of flat tires \\
	 &  & \textbullet\ Prone to engine failure \\
	 &&	\textbullet\ May use highway\\
	Donald & Car & \textbullet\ Risk of flat tires \\
	&  & \textbullet\ Prone to engine failure \\
	&&	\textbullet\ May use highway\\
	\bottomrule  
\end{tabular}
\end{table}

\subsection{Event squares}\label{ssec:randomeventsquares}
The board hosts 26 event squares. Landing on one of these squares usually entails skipping one or more turns. Aside from the shortcuts, only Square $[71]$ is seemingly a positive event, but it moves the player to Square $[74]$, where an event card should be picked up. The events are described in Table~\ref{table:descriptioneventsquares}. Note that event squares are largely independent of transport mode.


\begin{table}[H]
	\caption{Description of event squares.}
	\label{table:descriptioneventsquares}
		\small
\begin{tabular}{ l  l l l}
		\toprule
		\textbf{Square(s)} & \textbf{Description} & \textbf{Effect} & \textbf{Affected transport modes} \\
		\midrule
	$[9]$  & `Drink a coffee' & Skip 1 turn&	All\\
	$[13]$  & `Junk on the road' & Skip 1 turn&	All\\
	$[17]$-$[19]$  & `Forbidden to take over' & Skip 1 turn&	All\\	
	$[24]$  & `Stop for picnic' & Skip 1 turn&	All\\
	$[29]$  & `Change currency' & Skip 2 turns&	All\\	
	$[39]$-$[41]$  & `Dangerous curb, slow down' & Skip 1 turn&	All\\
	$[45]$  & `Bicycle lane' & May use shortcut& Bicycle, Walking\\
    $[50]$  & `Take a break' & Skip 2 turns&	All\\
    $[56]$  & `Fill up tank at gas station' & Skip 1 turn& Car, Motorcycle, Bus\\	
    $[63]$-$[65]$  & `Slow down' & Skip 1 turn& All\\
    $[71]$  & `Chased away from money bin' & Move forward 3 squares & All\\
    $[81]$  & `Nice spot, take a picture' & Skip 1 turn& All\\
    $[83]$  & `Nauseous, headache.  & Skip 1 turn & All\\
              & Go to first aid' &&\\
    $[90]$          & `Eat a bite' &Skip 1 turn & All\\ 
     $[91]$          & `Have a drink' &Skip 1 turn & All\\
   	$[92]$  & `No money? Wash dishes  & Skip turns, throw 6&	All\\		
          & to pay the bill' &to proceed&\\
	$[98]$  & `Lost in dark tunnel' & Skip turns, throw 2  &	All\\
      	  &&to proceed to Square 99&\\
      	  $[100]$  & `Highway' & May use shortcut& Car, Motorcycle, Bus\\
    $[105]$  & `Adhere to speed limit' & Skip 1 turn&	Car, Motorcycle, Bus\\
    $[112]$  & `Forgot camping card' & Move to Square 0&	All\\
\bottomrule
\end{tabular}
\end{table}

\subsection{Random event cards}\label{ssec:randomeventcards}
There are 11 random event cards and 17 board squares that require drawing such a card. Before starting the game, the event cards are shuffled before being stacked. Once drawn, the card is placed on the bottom of the stack again, so their order remains fixed. Compared to the event squares, the cards more frequently prescribe board movements rather than skipping turns. The random event cards -- denoted by \# -- are described in Table~\ref{table:randomeventcards}.

\begin{table}[H]
	\caption{Description of the event cards.}
	\label{table:randomeventcards}
	\small
\begin{tabular}{ l  l l l}
	\toprule
	\textbf{Card} & \textbf{Description} & \textbf{Effect} & \textbf{Affected transport modes} \\
	\midrule			
	$\#1$ & `Headwind, move backwards' & Throw die again to move backward & Bicycle, Motorcycle, Walking \\
	$\#2$ & `Tailwind, move forward' & Throw die again to move forward & Bicycle, Motorcycle, Walking \\
	$\#3$  & `Walker has blister' & Skip 1 turn&	Walking\\
	$\#4$  & `Walker gets a ride' & Move to next random event square  &	Walking\\
	$\#5$  & `Flat tire' &Skip 1 turn& Bus, Bicycle, Car, Motorcycle\\
$\#	6$ & `Work in progress: & Skip 3 turns & All\\
	 &  road maintenance' &  & \\
	$\#7$  & `Forgot route map at home' & Move to Square $[0]$&	All\\
	$\#8$  & `Forgot photo camera  & Move to Square $[37]$&	All\\
	& at saloon'&(when at Square $[26]$ or higher)&\\
	$\#9$  & `Bought postcards, put  & Move to Square $[32]$ &	All\\
	&them in the mailbox'&(from any current square)&\\
$\#10$  & `Engine failure, wait for  & Skip 2 turns& Bus, Car, Motorcycle\\
&roadside assistance'&&\\
$\#11$  & `Heavy rainfall' & \textit{All} affected transport modes&  Bicycle, Motorcycle, Walking\\
  &  & move back 3 squares&  \\
	\bottomrule  
\end{tabular}
\end{table}

\section{Analysis}\label{sec:analysis}
To draw statistically significant conclusions, we repeat perform trials of 10,000 game repetitions for each scenario that we test. We create scenarios ranging from 2 to 5 players; the five-player game is our base scenario. The game is coded in Python 3.7, which considerably speeds up the proceedings when compared to manually performed repetitions. A basic Monte Carlo simulation is at the heart of the model (see \citet{raychaudhuri2008} for an introduction on the technique).

In Section~\ref{ssec:gameproperties}, we discuss various gaming properties. An enjoyable GG should score well on various measurable properties; serious deviations from the target values may indicate that there are structural problems with the game. We draw upon the drama criteria defined by \citet{neto2016}, making some adjustments and additions on the way. In particular, we focus on the game duration and the balance between characters.

\subsection{Game properties}\label{ssec:gameproperties}

\begin{figure}
	\subfloat[2 players (average = 123, st. dev = 91.6)]{\includegraphics[width = 3in]{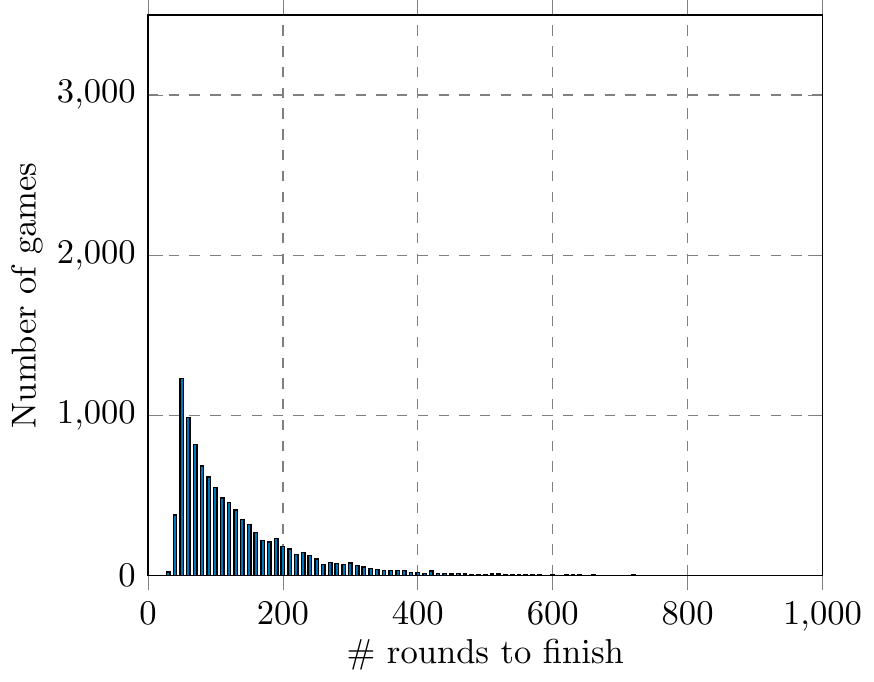}} 
	\subfloat[3 players (average = 92, st. dev = 60.3)]{\includegraphics[width = 3in]{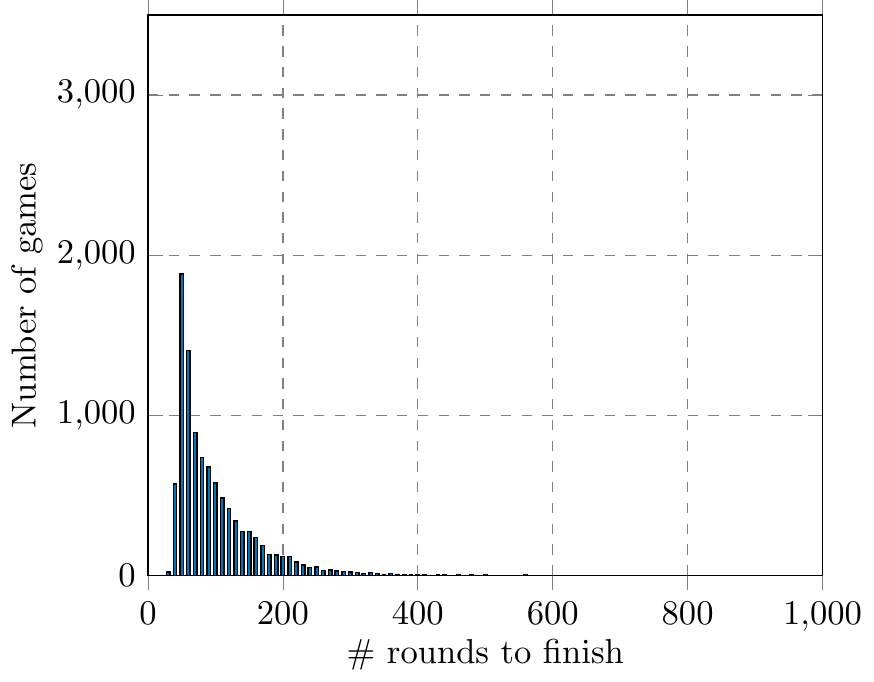}}\\
	\subfloat[4 players (average = 78, st. dev = 46.7)]{\includegraphics[width = 3in]{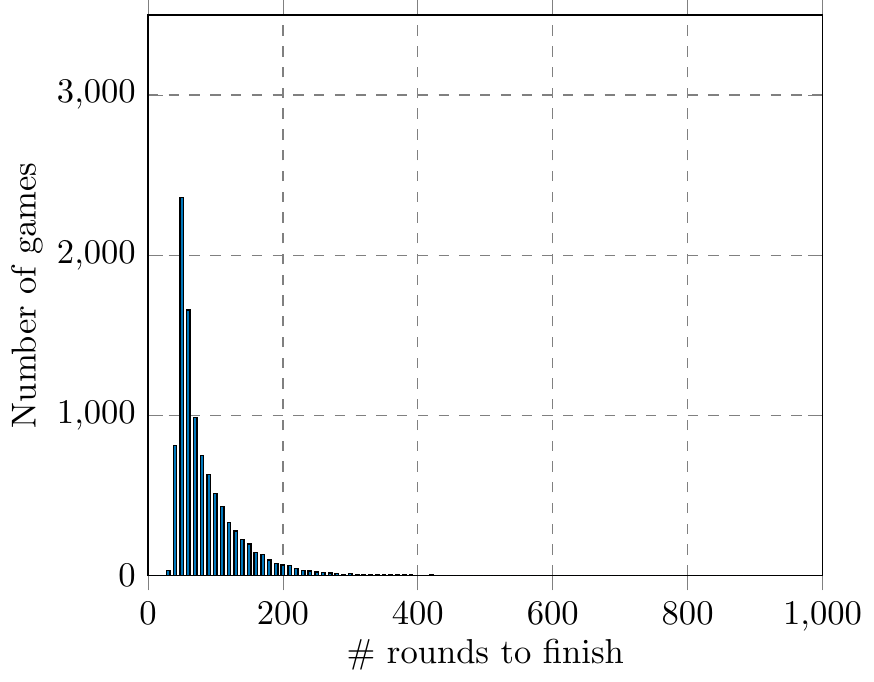}}
	\subfloat[5 players (average = 69, st. dev = 35.9)]{\includegraphics[width = 3in]{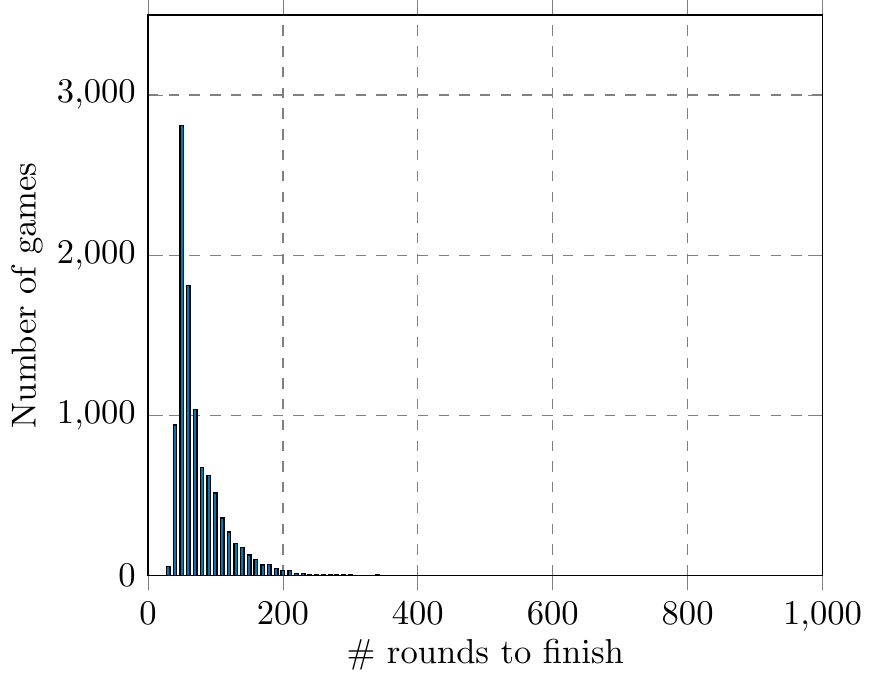}} 
	\caption{Histograms showing the simulated numbers of rounds to finish a game, for 2-5 players.}
	\label{fig:histogram}
\end{figure}

For each trial, Figure~\ref{fig:histogram} shows the number of rounds it takes the winner to complete the game. Table~\ref{table:rounds} summarizes the statistical properties. The average number of rounds ranges between 69 (for 5 players) and 123 (for 2 players). Although fewer rounds are required when adding more players, more die throws are needed (from 246 up to 345). As observed by \citet{neto2016}, the game durations roughly appear to be lognormally distributed. With fewer players, the right-hand tail packs relatively more mass. As indicated by the tails, it is not uncommon that games take much longer than indicated by the averages. We therefore perform some additional analysis and find that in the highest quartile averages are between 119 and 251 rounds, showing that lengthy games are fairly common.

\begin{table}[H]
	\caption{Statistical summary of the duration of the game, expressed in numbers of rounds.}
	\label{table:rounds}
	\small
	\begin{tabular}{ l  l l l l}
		\toprule
		\textbf{Metric} & \textbf{2p} & \textbf{3p} & \textbf{4p} & \textbf{5p} \\
		\midrule			
		Mean & 123 & 92 & 78 & 69\\
		Median & 92 & 72 & 61 & 55\\
		St. dev. & 92 & 60 & 47 & 36\\
		Mean Q1 & 46 & 43 & 41 & 40\\
		Mean Q2 & 75 & 59 & 53 & 50\\
		Mean Q3 & 122 & 91 & 76 & 66\\
		Mean Q4 & 251 & 176 & 143 & 119\\
		Min & 23 & 24 & 24 & 20\\
		Max & 878 & 625 & 416 & 386\\
		\bottomrule  
	\end{tabular}
\end{table}

\begin{figure}[H]
	\subfloat[Example of two-player game, taking 386 rounds to complete (7 returns to start).]{\includegraphics[width = 3.25in]{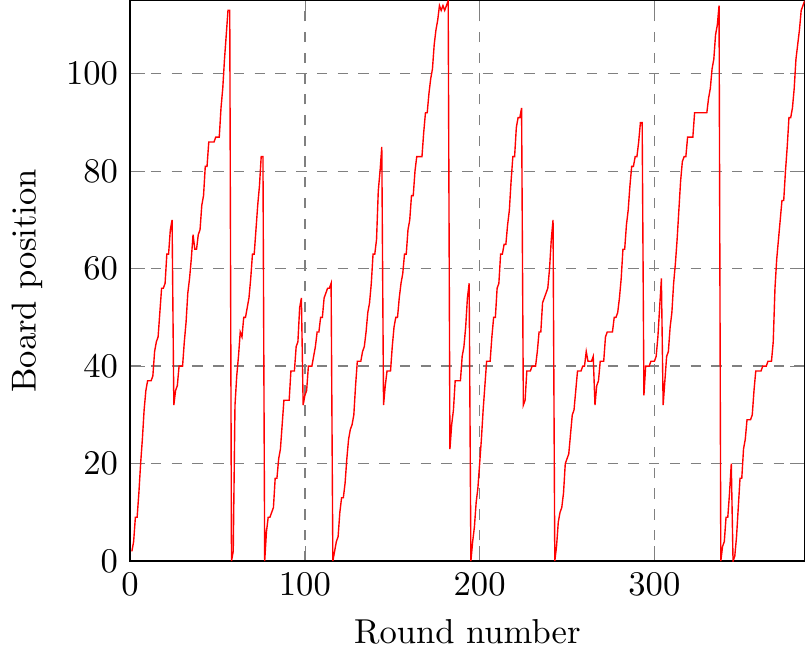}}
	\subfloat[Example of five-player game, taking 82 rounds to complete (1 return to start).]{\includegraphics[width = 3.25in]{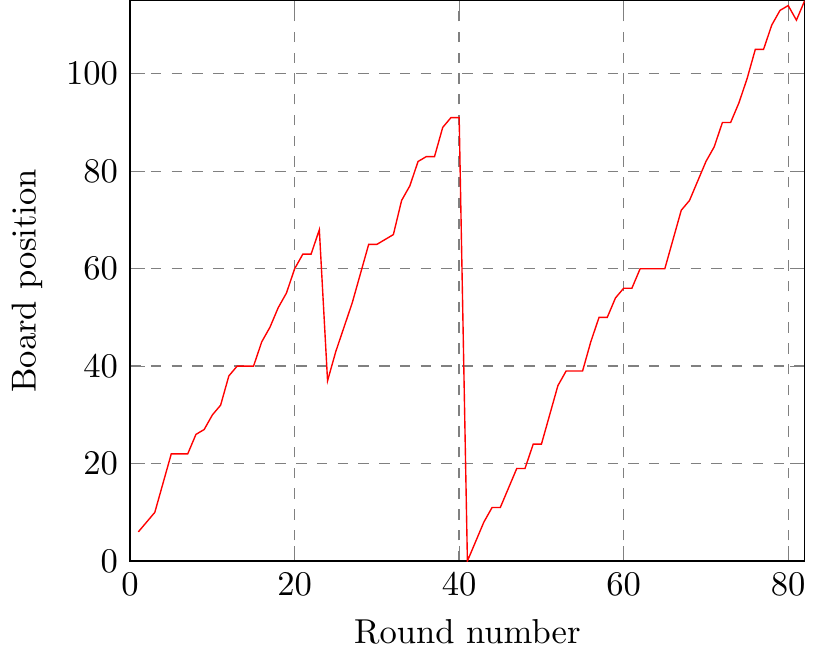}}
	\caption{Sample paths of winner's progression across the board.} \label{fig:examplepath}
\end{figure}

We conservatively estimate that in manual play, each turn takes 10 seconds to execute. In case of an event, this may well be longer, not in the least due to the emotional processing of setbacks. Also recall that a player may need to advance through multiple events within a single turn. Given our time assumption, on average the game can be completed within an hour, ranging between 41 minutes for the two-player game and 58 minutes for the five-player game. This seems to be an acceptable duration when played during long summer holidays. We find that 25.4\% of the games last over 1 hour and 3.7\% last over two hours; in these instances the game might start to feel tedious. Figure~\ref{fig:examplepath} illustrates the progression of the winning player across the board for a normal gameplay and a gameplay from the highest quartile (not a negligible outlier). Especially in two-player games, players are frequently forced to return to start, which might make for a somewhat monotonous experience.

\begin{figure}[H]
	\includegraphics{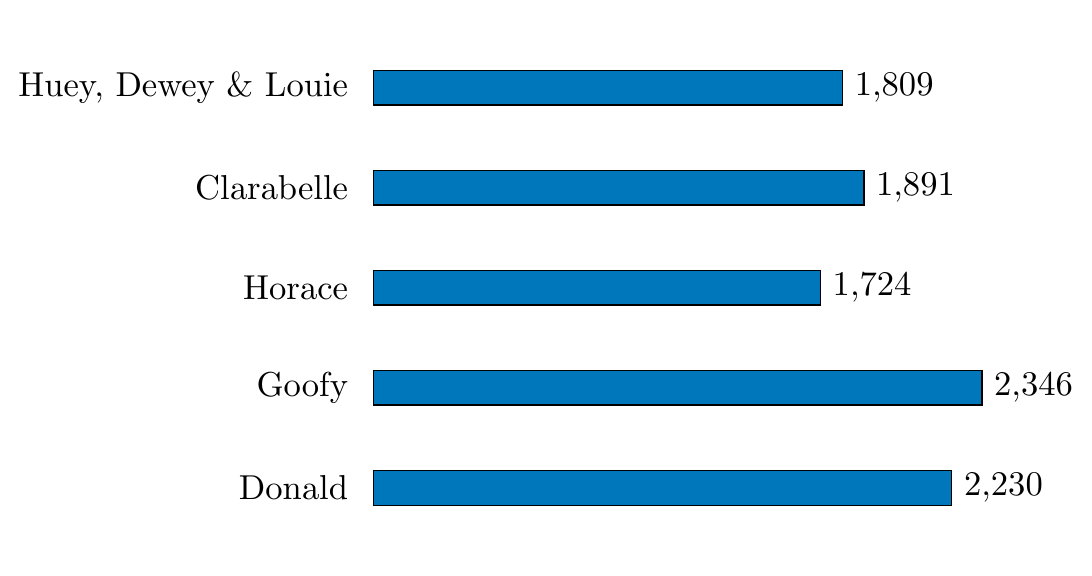}
	\caption{Number of wins per character, measured over 10,000 five-player games. The characters Goofy and Donald significantly outperform their competitors.} \label{fig:characterperformance}
\end{figure}

We proceed by measuring the win criteria, which indicate how balanced the game is. We evaluate whether (i) the character affects the probability of winning and (ii) the starting position affects the probability of winning. If, on average, each player has an equal chance to win, there is more tension. First, we address the character dimension. The characters may have different attributes that influence their game progress, but you would expect each player to have an equal chance to win. For the five-player game trial, the number of wins is shown in Figure~\ref{fig:characterperformance}. The characters Goofy and Donald (recall they are identical in terms of properties) significantly outperform the others -- as verified by a $z$-test at the 99\% level -- being roughly 30\% more likely to win. Thus, in this regard the game is not well-balanced. Tests on subsets of players (i.e., 2 to 4 players) do not show any substantial deviations; Goofy and Donald remain most likely to win in all configurations. Second, we evaluate the relationship between the starting position (remind that the player throwing 6 first may start) and chance of winning. We find no link whatsoever between the starting position and the number of wins; thus, in this aspect the game is balanced.

Next, we discuss the lead criteria, comprising (i) the average number of turns the winner has been in the lead before securing the win and (ii) the number of leaders at some point during the game. We start with the first criterion. If a player typically leads for many consecutive turns before securing the win, the game apparently lacks tension; the leading player is expected to win. Therefore, a low lead value makes for more dramatic games. However, a very low value implies that leading has no value at all, whereas there should be a certain level of anticipation involved with being in the lead. On average, the winner in the DDHG leads for 13 rounds, translating to $\sim$19\% of the game duration. In our opinion, this is a reasonable value. Regarding the second criterion: if a game typically only has one player in the lead, there is little chance to catch up once fallen behind. On the other hand, if every player expects to lead at some point, this might indicate a lack of structure in the game. According to \citet{neto2016}, an ideal value should be somewhere in the middle. For the DDHG, the average number of leaders during a game are 1.98 (2 players), 2.90 (3 players), 2.70 (4 players) and 4.40 (5 players); i.e., virtually ever player leads at a given time. For this reason, being in the lead does not feel like a decisive factor. However, for larger GG boards it might be natural that many players lead; it does not feel obstructive in this setting. In combination, the lead criteria indicate that early in the game there may be many shifts in the lead position, but in the final stages leading becomes more relevant. This seems to be a healthy setup.

We proceed with a criterion on idleness. Events that require the player to skip one or more turns are an important component of GGs, as they keep the player sidelined while their competitors catch up or extend their advantage. However, too much inactivity might be detrimental to the gaming experience. In the DDHG, the average percentage of turns waiting (relative to the number of rounds played) ranges from 10\% (in two-player games) to 26\% (in five-player games) turns. For the four-player and five-player games, we would argue that the time spent idly is on the high side, being sidelined for (almost) a quarter of the game as a passive observer.

The main findings are summarized below:

\begin{itemize}
	\item {Some characters are considerably more likely to win than others;}
	\item {The average game duration is acceptable, but games in the highest quadrant take too long to complete;}
	\item {Most players lead at some point, but leading towards the end of the game is a relevant predictor for securing the win;}
	\item  {Idleness in games with four or five players rises quite high, requiring players to skip about a quarter of all turns.}
\end{itemize}

\subsection{Narrative}
The qualitative aspect that we touch upon in this analysis is the importance of the narrative. Following the work of \citet{jenkins2004}, the events might be defined as `micro-narratives'; small story elements that evoke some emotional response. In this sense, the DDHG is richer than the traditional GG. Even though many events have the same effect (e.g., skip 1 turn), the varying micro-narratives keep the game enjoyable for a longer time. Furthermore, the unique and recognizable characters create a stronger and more prolonged emotional involvement, when comparing role-playing GGs to the traditional version.

The downside of these varying micro-narratives is that repetition makes them less credible. Event card $\#7$ (`forgot route map') and event square $[112]$ (`forgot camping card') have a major impact on progress, as they send you all the way back to start. When such events occur to the same player multiple times -- as illustrated in Figure~\ref{fig:examplepath} -- this undermines the credibility of the game; the same micro-narrative is repeated. For example, players feel they already `collected' the map at some point; returning home to collect the map twice feels less plausible than `forgetting' both the map and the camping card once. Especially when playing with fewer characters, players are bound to experience identical micro-narratives more often. This may unnecessarily detriment the gaming experience; allowing certain high-impact events to occur only once to each player would both avoid games of excessive duration and improve the quality of the overarching narrative.

\subsection{Analysis on balancedness}
As indicated before, the characters Goofy and Donald are significantly more likely to win than the others. This section addresses potential causes for this and suggest how to level the playing field; the key comparisons are summarized in Table~\ref{table:balance}. When measuring the number of event squares visited and event cards drawn, we count only the ones influencing the character, e.g., a flat tire for a walking character is not counted. 

Goofy and Donald travel by bus and car respectively; they may use the highway shortcut, are not affected by events that weather, but are prone to flat tires and engine failures. On average, Goofy and Donald skip 0.9 turns more than Clarabelle, 0.6 more than Huey, Dewey \& Louie, but 0.7 less than Horace. Event squares $[50]$ and $[105]$ only apply to motorized vehicles, which explains why they are slightly more likely to skip a turn. Despite this, they still perform better than others, so their performance is not explained by the number of turns skipped.

Another potential reasons are the shortcuts, which differ for the motorized and non-motorized transport modes. However, on average the bicycle lane is traversed more often than the highway (0.7 to 0.2). This is partially due to voluntary detours (to avoid landing on Square $[112]$), but mainly because the highway is located considerably closer to the end; the early stages of the board are traversed much more frequently. In any case, the shortcuts do not favor the motorized vehicles.

When looking at the number of event cards drawn, we find that the number is significantly lower for Goofy and Donald than for the other characters (5.0 to 6.3-7.9). This appears to explain the root cause of the unbalance. Walkers, bicycles and motorcycles are affected by the wind (two event cards) and rain (affecting all players whenever the card is drawn), while walkers may also hitchhike (card \#4, requiring them to draw another card), such that on average they draw more cards than the characters traveling by car or bus. As most cards have negative effects, this explains the relatively poorer performance. We find that, on average, Goofy and Donald are set back 14 to 24 squares less due to random events than the other characters, explaining their competitive advantage.

\begin{table}[H]
	\caption{Numerical attributes explaining the balance within the game.}
	\label{table:balance}
	\small
	\begin{tabular}{ l  l l l l l}
		\toprule
		\textbf{Character} & \textbf{\# event} & \textbf{\# event} & \textbf{\# turns} & \textbf{\# shortcuts} & \textbf{\# squares} \\
		 & \textbf{squares} & \textbf{cards} & \textbf{waiting} &  & \textbf{moved} \\
		\midrule			
		Huey, Dewey \& Louie  & 12.4 & 7.9 & 17.3 & 0.7 & -93\\
		Clarabelle & 12.5 & 6.3 & 17.0 & 0.7& -89\\
		Horace & 12.8 & 6.8 & 18.6 & 0.2&-83\\
		Goofy & 12.3 & 5.0 & 17.9 & 0.2&-68\\
		Donald & 12.3 & 5.0 & 17.9 & 0.2&-69\\
		\bottomrule  
	\end{tabular}
\end{table}

We test various modifications of the original DDHG; Table~\ref{table:revisedgame} shows the results of these alternative game designs. Event card $\#11$ seemingly contributes most to the unbalance (`heavy rainfall: bicycle, motorcycle and walkers move back 3 squares'). Although its effect is seemingly minor, it affects the characters each time \textit{any} of the players draws the card, on average four times per game. Furthermore, these extra moves increase the risk of having to draw another event card.  We show that removing card $\#11$ makes for a more level game, although it somewhat favors the Clarabelle character. Removing cards $\#1$ and $\#2$ (headwind and tailwind) also reduces the performance gap to some extent, but not sufficiently to negate the differences between characters. We find that removing cards $\#2$ and $\#11$ from the deck yields a fairly balanced game without defining new events.

\begin{table}[H]
	\caption{Number of wins per character for several game variant that omit certain event cards.}
	\label{table:revisedgame}
	\small
	\begin{tabular}{ l  l l l l l}
		\toprule
		\textbf{Character} & \textbf{Original} & \textbf{No rain} & \textbf{No wind} & \textbf{No rain} & \textbf{No rain and}\\
		& &  &  & \textbf{and wind} & \textbf{tailwind}\\
		\midrule			
		Huey, Dewey \& Louie  & 1,809 & 1,953 & 1,870 & 1,997&1,977\\
		Clarabelle & 1,891 & 2,184 & 1,908 & 2,128&2,066\\
		Horace & 1,723 & 1,911 & 1,791 & 1,893&1,901\\
		Goofy & 2,345 & 1,949 & 2,196 & 1,979&2,015\\
		Donald & 2,230 & 2,003 & 2,235 & 2,003&2,036\\
		\bottomrule  
	\end{tabular}
\end{table}

\begin{figure}[H]
	\subfloat[2 players (average = 85, st. dev = 34.1)]{\includegraphics[width = 3in]{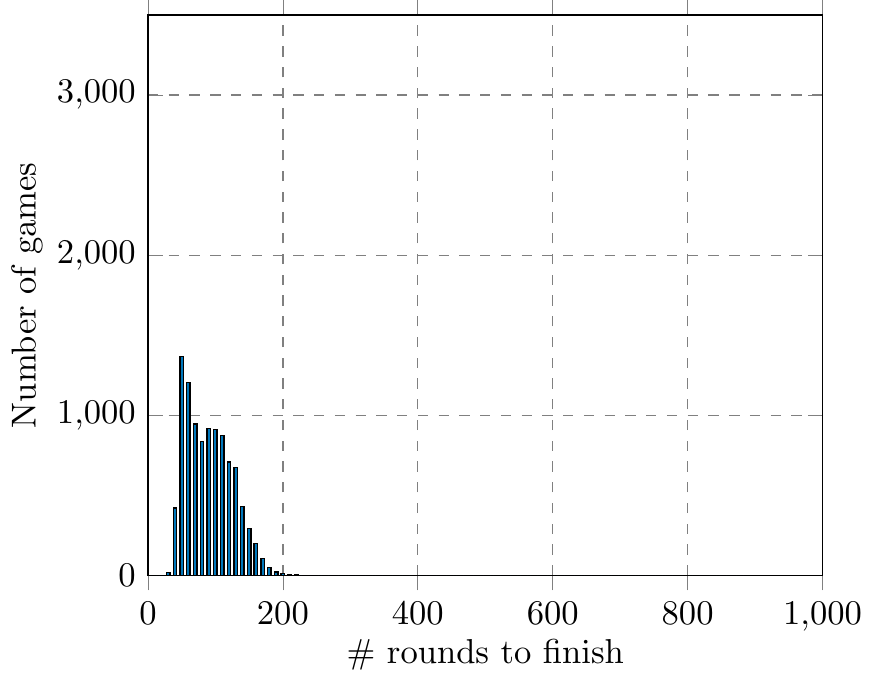}} 
	\subfloat[5 players (average = 58, st. dev = 21.2)]{\includegraphics[width = 3in]{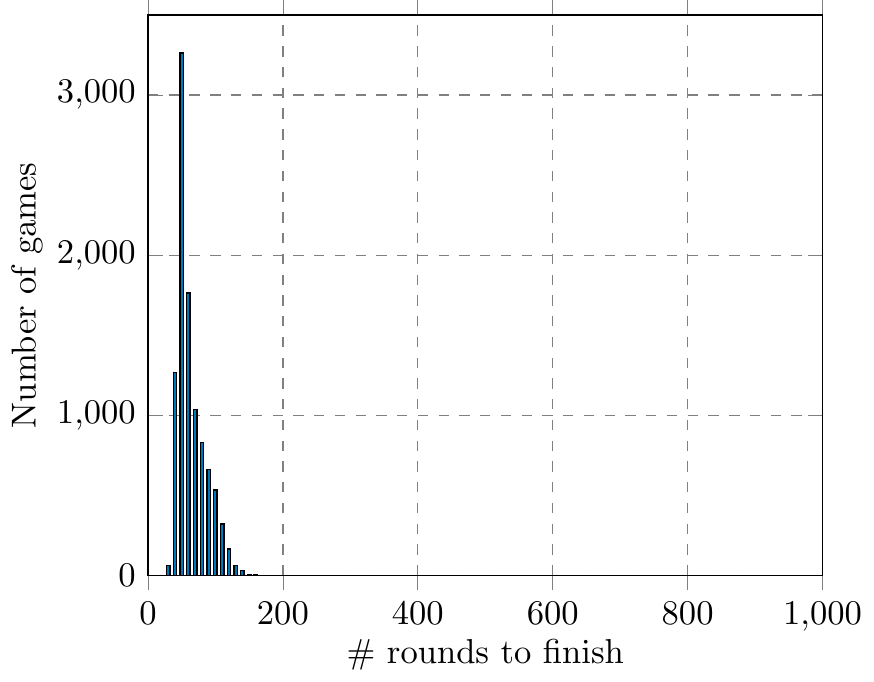}}
	\caption{Histograms showing the simulated numbers of rounds to finish the revised game, for 2 and 5 players.}
	\label{fig:histogramrevised}
\end{figure}

Finally, circling back to the narrative part, we place a cap on the route map- and camping card events, enforcing that both can only be `forgotten' once by each character. Thus, a player can only be sent back to start twice during a game.  Combined with removing the effects of rain and tailwind, this makes for a more balanced game with lower duration, especially cutting out the extremes. For two-player games, we find an average number of turns of 85 (was 123) and a maximum of 237 (was 878). For five player-games, we find an average number of turns of 58 (was 69) and a maximum of 153 (was 386). The histograms of the revised game are shown in Figure~\ref{fig:histogramrevised}. Although it would be quite a stretch to call these histograms normally distributed, they are notably more symmetrical than the original ones. For future studies, it would be interesting to compare the enjoyability of board games with normal- and lognormal distributions of duration.

\section{Conclusions}\label{sec:conclusion}
In this paper, we developed and applied a Monte Carlo simulation model to evaluate a Game-of-the-Goose role-playing variant. We evaluated this game on a number of metrics that indicate how enjoyable the game is. We found that the game duration at times is a cause for concern. The average duration seems acceptable, but excesses occur too frequently. When considering the highest quartile, games last well over an hour, at which point the game may test the resilience of the intended audience. Another major conclusion is that the game characters are not fairly balanced: the characters traveling by bus or car have roughly 30\% more chance to win. Unfortunately, this is not an encouraging outcome from the perspective of the sustainable transport domain, somewhat undermining the author's findings in earlier research \citep{vanheeswijk2017,vanheeswijk2018,vanheeswijk2019,vanheeswijk2020}. Without the harsh influence of pouring rain and blustery winds, however, the future of sustainable transport would look much brighter, giving characters afoot and on bicycle a fair chance of winning. Finally, we argue that the richer narrative in role-playing games aids to increase the longevity of the game, but also that repeated micro-narratives -- especially high-impact ones -- have an adverse effect on enjoyability.

In line with \citet{neto2016}, this paper provides an example of how a simple Monte Carlo simulation may be used to analyze key properties of the game. Such an approach could aid board game designers in an effort to improve the balance of their games; especially in role-playing games, it may be difficult to foresee how the differences between characters pan out. Tracking relevant metrics during repeated gameplay helps to identify possible caveats and test modifications, which will ultimately lead to more balanced and enjoyable games.

\bibliographystyle{apalike}
\bibliography{bibliographyDDHG}

\end{document}